\documentclass[11pt]{article}
\usepackage[normalem]{ulem}
\pdfoutput=1 

\usepackage{jheppub} 
\usepackage{url}
\usepackage{caption}
\usepackage{subcaption}
\usepackage{extarrows}

\usepackage[latin9]{inputenc}
\usepackage{float}
\usepackage{amsmath}
\usepackage{amssymb}
\usepackage{graphicx}
\usepackage{esint}
\usepackage{hyperref}
\usepackage{comment}
\usepackage{color}
\usepackage{microtype}
\usepackage{cleveref}
\usepackage{breakurl}
\usepackage{bbm}
\newcommand{\be}{\begin{equation}}
\newcommand{\ee}{\end{equation}}
\newcommand{\ben}{\begin{displaymath}}
\newcommand{\een}{\end{displaymath}}
\newcommand{\bea}{\begin{eqnarray}}
\newcommand{\eea}{\end{eqnarray}}
\def\K{K{\"a}hler }
   \newcommand{\rf}[1]{(\ref{#1})}
\newcommand{\vp}{\varphi}

\def\be{\begin{equation}}
\def\ee{\end{equation}}
\def\bea{\begin{eqnarray}}
\def\eea{\end{eqnarray}}
\def\ba{\begin{array}}
\def\ea{\end{array}}
\def\bit{\begin{itemize}}
\def\eit{\end{itemize}}

\def\a{\alpha}

\def\vp{\varphi}

 \makeatletter

\allowdisplaybreaks

\makeatother

\makeatletter
\DeclareRobustCommand{\rcite}[1]{%
  \rcite@aux#1,\@nil{#1}%
}
\def\rcite@aux#1,#2\@nil#3{%
  \if\relax#2\relax
    Ref.~\cite{#3}%
  \else
    Refs.~\cite{#3}%
  \fi
}
\makeatother

\hypersetup{
    colorlinks = true,
    citecolor = {blue},
    linkcolor = {blue},
    urlcolor = {blue},
}

 \title{\rm {\bf \LARGE \boldmath  {Double Exponents  in $SL(2,\mathbb{Z})$   Cosmology}}}

\author{Renata Kallosh\, and }
\author{Andrei Linde}

\affiliation{Stanford Institute for Theoretical Physics and Department of Physics,\\ Stanford University, Stanford, CA 94305, USA}
\emailAdd{kallosh@stanford.edu}
\emailAdd{alinde@stanford.edu}

\abstract{  
$SL(2,\mathbb{Z})$ invariant $\alpha$-attractor models proposed in \cite{Kallosh:2024ymt} have plateau potentials with respect to the inflaton and axion fields. The slope of the potential in the inflaton direction is exponentially suppressed at large values of the inflaton field,  but the slope of the potential in the axion direction is {\it double-exponentially suppressed}. Therefore, the axion field remains nearly massless and practically does not change during inflation. The inflationary trajectory in such models is stable with respect to quantum fluctuations of the axion field. We show that isocurvature perturbations do not feed into the curvature perturbations during inflation, and discuss the possibility of such transfer at the post-inflationary stage.
}

\begin{document}

\maketitle


\parskip 6pt

\section{Introduction}\label{Sec:intro}

In this paper, we continue investigation of the cosmological $\alpha$-attractor models with $SL(2,\mathbb{Z})$ symmetry proposed in \cite{Kallosh:2024ymt,Casas:2024jbw,Schimmrigk:2021tlv}, where the potentials depend on $SL(2,\mathbb{Z})$ invariants. 

We  focus on a class of potentials introduced in \cite{Kallosh:2024ymt} depending on Klein's   $j$-function
\be
\left( {\cal L} (\tau, \bar \tau)\over \sqrt{-g}\right)^{SL(2, \mathbb{Z})} =  {R\over 2} + {3\alpha\over 4} \, {\partial \tau \partial \bar \tau\over ({\rm Im}  \tau )^2}- V\Big (j (\tau) ,  \overline { j (\tau) } \Big )   .
\label{hyper}\ee
All other $SL(2,\mathbb{Z})$ invariant potentials in \cite{Kallosh:2024ymt,Casas:2024jbw} have analogous features with regard to the axion  $\theta$ and the inflaton $\vp$, where
\be
\tau = x+iy=\theta + i e^{\sqrt{2\over 3\alpha} \vp}  \ .
\label{coo}\ee
Three features of our cosmological models play an important role in the analysis.
\begin{enumerate}
  \item {\it The geodesics in hyperbolic geometry in half-plane coordinates \rf{coo} are
straight vertical lines} 
\be
\theta = {\rm const}
\ee as well as certain semi-circles, see Fig. 1 and the discussion in the Appendix in \cite{Kallosh:2024pat}.  The reason for the straight line geodesics in curved space (line of the shortest distance) is the property  of the hyperbolic geometry that $\Gamma^\theta_{\vp\vp}=0$ and the geodesic equation 
\be
\theta^{''} -2\sqrt{\frac2{3\alpha}}\vp'\theta'=0 \ .
\ee
In general geometry, this equation would have a term $\Gamma^\theta_{\vp\vp} \vp' \vp'$, which would prevent a straight line solution for the geodesics $\theta'=0$.

\item {\it The potential at large positive values of the inflaton field $\vp$ has the $\alpha$-attractor behavior,}
\be
V(\vp,\theta)  \propto  1- e^{-\sqrt{2\over 3\alpha} \varphi} +...  .
\ee

\item {\it The derivatives of the potential with respect to the axion field are double-exponentially suppressed during inflation}
\be \label{supprV} 
{\partial V\over \partial \theta}  \propto   e^{-2 \pi e^{\sqrt{2\over 3\alpha} \varphi}} \sin(2\pi \theta)\, .
\ee
This feature makes the geodesic  $\theta =$ const a correct solution even with an account of the potential, since the contribution from the potential in the $\theta$ equation during inflation can be neglected.
\end{enumerate} 
The number of e-foldings of inflation beginning at some value of the inflaton field $\vp$ in $\alpha$-attractor models  is given by
\be\label{ne}
N_{e} = {3\alpha \over 8} e^{\sqrt{2\over 3\alpha} \varphi} \ .
\ee
Therefore the double-exponential factor in \rf{supprV} can also be represented as 
\be\label{gogsup}
e^{-2 \pi e^{\sqrt{2\over 3\alpha} \varphi}} = e^{-{16\pi N_{e}\over 3\alpha}} \sim 10^{-1200}  = {\rm googol}^{{-12}}
\ee
for $\alpha = 1/3$ and $N_{e} \sim 55$. Thus the double exponential behavior of the axion potential during inflation
brings in the googol scale into cosmology.
One  googol\footnote{Google, the name of the search engine and the verb that refers to searching the Internet using the Google search engine are plays on the mathematical expression googol $=10^{100}$.} equals $10^{100}$ and googol$^{-12}= 10^{-1200}$. These are not the numbers that we often encounter in physics. Amazingly, these numbers appear in the description of derivatives of the $SL(2,\mathbb{Z})$ invariant potentials. The origin of the double-exponential suppression \rf{supprV} will be discussed in section \ref{InPot}.

It is instructive to compare inflation in the theory of $SL(2,\mathbb{Z})$ invariant $\alpha$-attractors \cite{Kallosh:2024ymt}  with inflation in closely related models in \cite{Achucarro:2017ing,Linde:2018hmx}, where the universality of a broad class of two-field $\alpha$-attractor predictions was established. It was found there that once inflation takes place near the boundary at $y \to 0$, the general inflationary predictions of T and E two-field models  coincide with general predictions of single-field $\alpha$-attractors for a large number of e-foldings $N_{e}$  \cite{Kallosh:2013yoa,Galante:2014ifa,Carrasco:2015uma,Kallosh:2021mnu}: 
\be \label{pred2}
 A_{s} = {V_{0}\, N_{e}^{2}\over 18 \pi^{2 }\alpha} \ , \qquad n_{s} = 1-{2\over N_{e}} \ , \qquad r = {12\alpha\over N^{2}_{e}}  \ .
\ee
Here $A_{s}$ describes the magnitude of perturbations, $n_{s}$ is the slope of their spectrum, and $r$ is the tensor-to-scalar ratio. 

It would be nice to use these results for investigation of the two-field evolution in the models with the $SL(2,\mathbb{Z})$ invariant plateau potentials \cite{Kallosh:2024ymt}. However, there is a substantial difference between these models and the models explored in \cite{Achucarro:2017ing,Linde:2018hmx}. The main difference is in the choice of the variable $y$ as a function of the canonically normalized inflaton field $\vp$. 

There are two such choices at large $\vp$
\be
y= e^{-\sqrt{2\over 3\alpha} \vp} \ ,   \label{lower}\ee
\be
y= e^{\sqrt{2\over 3\alpha} \vp} \ .  \label{upper}\ee
In the first case, implemented in the models of Refs.  \cite{Achucarro:2017ing,Linde:2018hmx}, $y$ becomes exponentially small at large $\vp$; in the second case, studied in  \cite{Kallosh:2024ymt}, $y$  it becomes exponentially large. 
Thus, large $\vp$ corresponds to inflation near the $y \to 0$ boundary as in eq. \rf{lower}. 

Meanwhile, inflation in the $SL(2,\mathbb{Z})$ invariant $\alpha$-attractors occurs in the limit  $y \to \infty $  as in eq. \rf{upper}. Therefore the results of 
 \cite{Achucarro:2017ing,Linde:2018hmx} do not automatically apply to the new models.

As we will show in this paper, the axion field also remains frozen during inflation in the theory of $SL(2,\mathbb{Z})$ invariant $\alpha$-attractors, but for a different reason.   At large values of the inflaton field $\vp$,  the slope of the potential in the axion direction practically vanishes. As a result, even if the axion field was moving initially, it eventually stops, and after that, the axion does not move at all until the field $\vp$ becomes small and inflation ends.

Thus, inflation happens just like in the single-field $\alpha$-attractors. However, one may wonder whether quantum fluctuations of the nearly massless axion field may destabilize the inflationary trajectory and/or affect the standard $\alpha$-attractor predictions. We will investigate these issues in the present paper.

\section{Inflationary potentials and power of double exponents}\label{InPot}

Cosmological $\alpha$-attractor models with $SL(2,\mathbb{Z})$ symmetry were developed only recently \cite{Kallosh:2024ymt,Casas:2024jbw,Schimmrigk:2021tlv}.  For definiteness, we focus here on potentials depending on Klein's Absolute Invariant $J(\tau)$ or  $j(\tau)$-function, where $j(\tau)=12^3  J(\tau) $ \cite{Kallosh:2024ymt}.
These modular functions are defined in the hyperbolic geometry of the upper half plane $\tau =x+iy$ where $ y>0$:
\be
j(\tau) = q^{-1} + \sum_{n=0} c_n q^n\, , \qquad c_0= 744  \ ,
\label{series}
\ee
where
\be
q= e^{2\pi i\tau}= e^{-2\pi y +2\pi i x}\qquad q^{-1}=e^{-2\pi i  \tau}= e^{2\pi y -2\pi i x} \ .
\label{q}\ee
Properties of the hyperbolic geometry in half-plain coordinates can be illustrated by Escher's picture of  Heaven and Hell, Fig. 1.

\begin{figure}[H]
\centering
\includegraphics[scale=0.5]{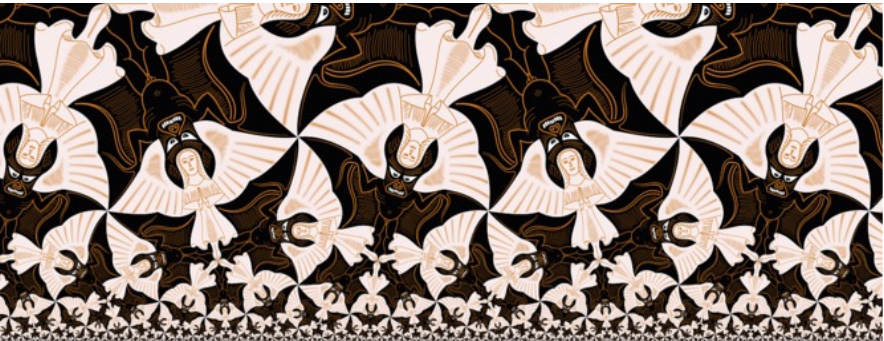}
\caption{\footnotesize Escher's picture of  Heaven and Hell in half-plane variables.
The boundary of the half-plane $y \to 0$ is the absolute, which cannot
be reached. The angels and devils look smaller and smaller near $y \to 0$. Inflation in $SL(2,\mathbb{Z})$ invariant potentials in \cite{Kallosh:2024ymt} takes place in the upper part of the half-plane close to  $y \to +\infty$. We presented and discussed this picture and the relevant cosmological $\alpha$-attractor models, mostly in disk variables in \cite{Kallosh:2015zsa}. }
\label{Escher}
\end{figure}
\begin{figure}[H]
\centering
\includegraphics[scale=0.12]{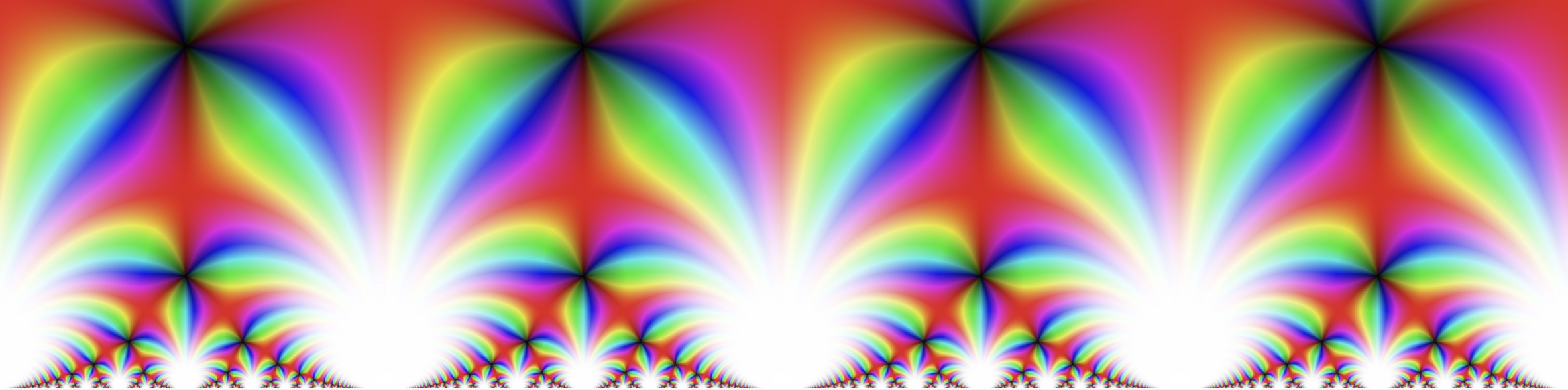}
\caption{\footnotesize    Fredrik Johansson's complex plot of the $j$-function, see fredrikj.net  in My blog: Modular forms in Arb. Here Arb is for a C library for arbitrary-precision ball arithmetic. }
\label{FJ}
\end{figure}
We have presented the complex plot of the $j$-function in Fig. \ref{FJ}.
At $y \gg 1$, the function $j(\tau)$ can be well represented by its first two terms:
\be
j(\tau)|_{y \gg 1}  = \Big( e^{2\pi y -2\pi i x} +  c_0  +\sum_{n=1} c_n e^{- 2n \pi y +2n \pi i x} \Big)|_{y\gg 1 }\to e^{2\pi y -2\pi i x} +   744  \ .
\label{largej}\ee
It is known from  \cite{Petersson:1932,Brisebarre:2005}  that  at  $n\geq 1$ 
\be
c_n\leq {e^{4\pi \sqrt{n}}\over \sqrt{2} \, n^{3/4}} \ ,
\ee
and therefore the terms of order $n$ as a function of $y$  are constrained to be 
\be
c_n q^n \leq {e^{4\pi \sqrt{n}}\over \sqrt{2} \, n^{3/4}} e^{-2\pi n y} \ .
\ee
Therefore, these series rapidly converge even for the values of $y$ close to the end of inflation. For example, already for $y=2$ 
the series converges very fast,  
\be\label{fast}
c_n q^n|_{y=2} \, \leq {e^{-4\pi (n - \sqrt{n})}\over \sqrt{2} \, n^{3/4}} \ .
\ee
Meanwhile, the  $j(\tau)$-function near the boundary $y \to 0$ is given by the series in eq. \rf{series} where all terms have to be taken into account. 
Nevertheless, one can investigate its global properties by using $SL(2,\mathbb{Z})$ invariance \cite{Kallosh:2024pat}.

\begin{figure}[H]
\centering
\includegraphics[scale=0.16]{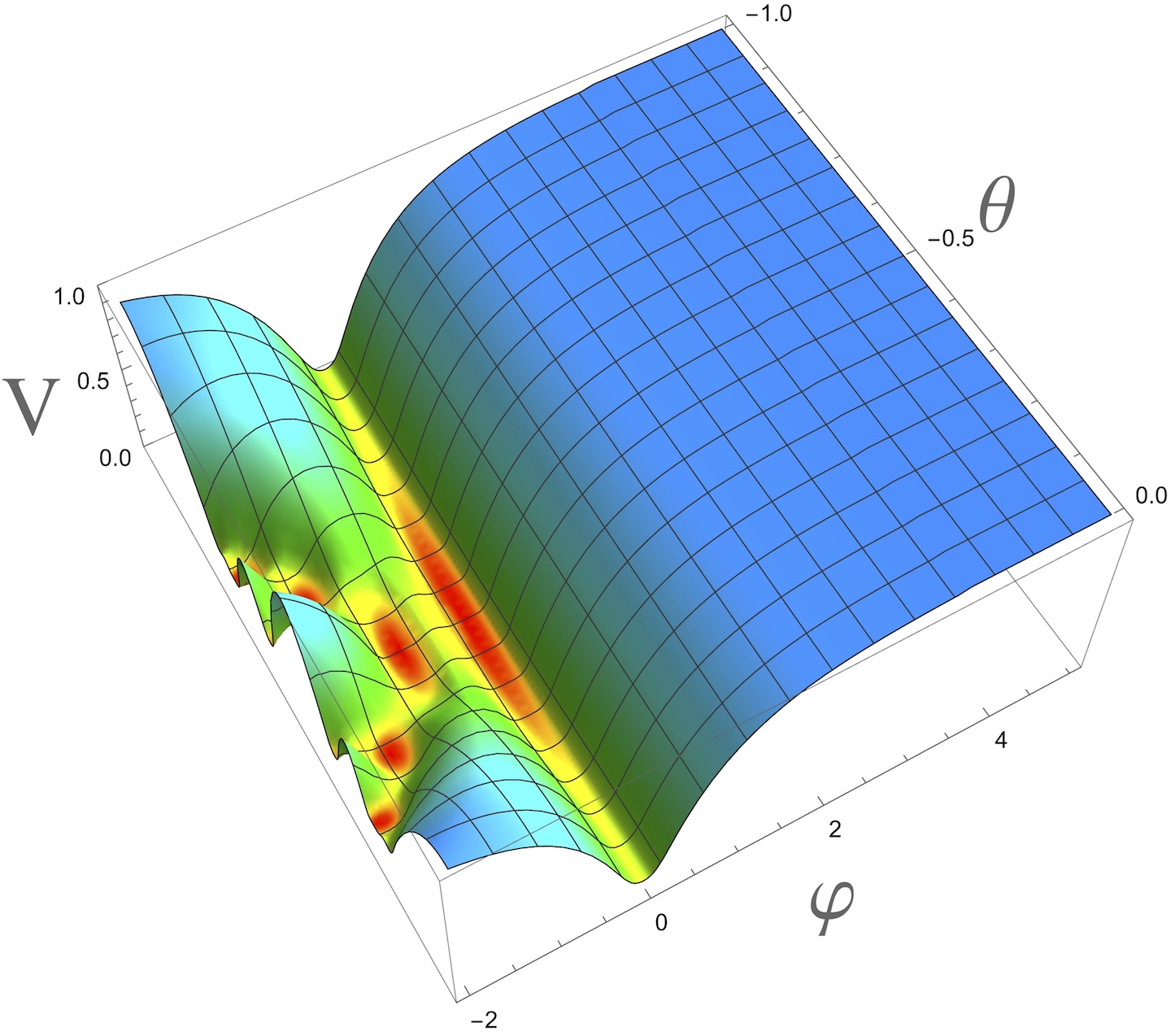}
\vskip -10pt
\caption{\footnotesize Potential \rf{Renata2},  for $\tau= \theta+ie^{\sqrt{2\over 3\alpha} \varphi}$ as a function of $\vp$ and $\theta$ in case   $3\alpha=1$.  }
\label{f2}
\end{figure}

As an example, consider the potential 
\be\label{Renata2}
V =V_0\Big (1-{\ln j(i)^{2}
\over  \ln \big(|j(\tau)|^2 + j(i)^{2}\big)}\Big ) \ .
\ee
Note that  $j[\tau=e^{2\pi i\over 3}]=0$ and the potential vanishes, $V(\tau=e^{2\pi i\over 3})=0$ and it has a minimum there. At $\tau =i$ the potential is positive and has a de Sitter saddle point.
We show the plot of this potential in Fig. \ref{f2}. As one can see, at large $\vp$ the potential is very flat in the $\theta$ direction.
We may also consider a slightly more general potential
\be
V =V_0\Big (1-{\ln \beta^{2}
\over  \ln \big(|j(\tau)|^2 + \beta^{2}\big)}\Big ) \ .
\label{beta}\ee
The potential in  \rf{Renata2} is a special choice of the one in \rf{beta} when $ \beta=j(i) = 12^3$. The so-called Klein invariant $J(\tau)= 12^{-3} j(\tau)$ at this point is $J(i)=1$.

 To understand the origin of the remarkable flatness of this potential in $\theta$ at large $\vp$, we will evaluate the derivative $V_\theta$ at large $\vp$. 
We can use the expansion in equation \rf{series} $j(\tau) = q^{-1} + \sum_{n=0} c_n q^n $, where    $q^{-1}= e^{2\pi (y -i \theta)}$ and $q= e^{-2\pi (y -i \theta)}$. At large $y$, as we argued in eqs. \rf{largej},   \rf{fast},
\be
j(\tau)|_{y\to \infty }  \to e^{2\pi y} e^{-i 2\pi \theta}  +744 \ ,
\label{large}\ee
where we have omitted all terms of the order $e^{-2\pi n y}$ with $n\geq 1$, so that
\be\label{j2a}
|j(\tau)|^2  = |e^{2\pi e^{\sqrt{2\over 3\alpha} \varphi}} +744\, e^{i 2\pi \theta}|^2 \ .
\ee
We can now calculate the derivative $V_\theta$ of the potential \rf{beta} during inflation.
 
As we already mentioned in the Introduction, during inflation in  $\alpha$-attractor models, we have $e^{\sqrt{2\over 3\alpha} \varphi }={8N_{e}\over 3\alpha}$, 
where $N_e$ is the number of e-folds of inflation beginning at $\vp = \varphi_{_{N_e}}$.  Therefore the exponential coefficient $2\pi e^{\sqrt{2\over 3\alpha} \varphi}$ in this equation can be represented as
\be
e^{2\pi e^{\sqrt{2\over 3\alpha} \varphi}} = e^{2\pi {{8N_{e}\over 3\alpha}}}  .
\ee
This term remains extraordinarily large all the way to the end of inflation with $N_{e} = 0$. Therefore one can keep only two leading terms in \rf{j2a}:
 \be
|j(\tau)|^2  =  e^{4\pi e^{\sqrt{2\over 3\alpha} \varphi}}\Big(1+ 1488 e^{-2\pi e^{\sqrt{2\over 3\alpha} \varphi}} \cos(2\pi \theta) \Big)\ .
\label{j2}\ee
As one can see, the $\theta$-dependent term is double-exponentially suppressed as compared to the first term. 

The double-exponential growth of $|j(\tau)|^2$ disappears when we calculate the potential \rf{beta}, but the double-exponential suppression of its $\theta$-dependence remains.
Using \rf{j2}, we find that the $\theta$-derivative  of the potential \rf{beta} during inflation is  
\be
{\partial V\over \partial \theta} =- 372\, V_{0 }\sin (2 \pi  \theta)\, \ln( \beta^2)  \  y^{-2}\,   e^{-2\pi y} \ .
\ee
Note that $V_\theta$ is multiplied by the exponentially large factor $\frac{2}{3\alpha}e^{2\sqrt{\frac{2}{3\alpha}}\vp}= \frac{2}{3\alpha} y^2$ in the equation for $\theta$ \rf{eqtheta}, so the relevant quantity is
\be
\frac{2}{3\alpha}y^{2} {\partial V\over \partial \theta}=- {248\, V_{0 }\over  \alpha}\, \sin (2 \pi  \theta)\, \ln(\beta^2)  \     e^{-2\pi y} \ .
\label{der}\ee
 The most important factor in this expression is the double-exponential factor  $e^{-2\pi y}=e^{-2\pi e^{\sqrt{2\over 3\alpha} \varphi}}$ which makes the last term in equation \rf{eqtheta} vanishingly small during inflation despite of the factor $y^2=\frac{2}{3\alpha}e^{2\sqrt{\frac{2}{3\alpha}}\vp}$.
 Replacing $y$ by $e^{\sqrt{2\over 3\alpha} \varphi}$ in eq. \rf{der} we find
\be
\frac{2}{3\alpha}y^{2} {\partial V\over \partial \theta} =- {248\, V_{0 }\over \alpha}\, \sin (2 \pi  \theta)\, \ln(\beta^2)  \      e^{-2\pi e^{\sqrt{2\over 3\alpha} \varphi}} \ .
\ee
The second derivative of the potential over $\theta$, $V_{\theta \theta}$ is also suppressed by the double exponential factor.

To understand the strength of this effect, we once again use the fact that during inflation in  $\alpha$-attractor models, $y = e^{\sqrt{2\over 3\alpha} \varphi }={8N_{e}\over 3\alpha}$. This means that
\be\label{slopetheta}
 \frac{2}{3\alpha}e^{2\sqrt{\frac{2}{3\alpha}}\vp} {\partial V\over \partial \theta}  =- {496\, V_{0} \over  \alpha^{2}}\, \sin (2 \pi  \theta)\, \ln(\beta^2)   \  e^{-{16\pi N_{e}\over 3\alpha}} \ .
\ee
As shown in the Introduction, the last coefficient in \rf{slopetheta} is extremely small. For example, for $N_{e} = 55$ and $\alpha = 1/3$, one finds that the derivative of the potential with respect to $\theta$, even multiplied by $e^{2\sqrt{\frac{2}{3\alpha}}\vp}$, as we need for eq. \rf{Es2a},  is suppressed by the factor $10^{-1200}$, see \rf{gogsup}.

This explains why the potential is incredibly flat in the axion direction during inflation. Note that the approach to the plateau in the inflaton direction is linear in $1/y\sim e^{-\sqrt{2\over 3\alpha} \varphi}$. The flatness is exponential for the inflaton but double exponential for the axion.

Now we will find the value of $N_{e}$ for which the maximal value of the right-hand side of  \rf{slopetheta}  becomes of the same order as the derivative of the potential with respect to $\vp$, which is given by
\be
V_{\vp}  = V_{0} \sqrt{2\over 3\alpha} e^{-\sqrt{2\over 3\alpha} \varphi} = V_{0} {8N_{e}\over 3\alpha}\sqrt{2\over 3\alpha}  \ .
\ee
One can easily check that, for $\alpha \lesssim 1$,  the term $V_{\vp} $ in the equation \rf{Es1a} remains much greater than the term $ \frac{2}{3\alpha}e^{2\sqrt{\frac{2}{3\alpha}}\vp} {\partial V\over \partial \theta}$ in  \rf{Es2a} all the way to the end of inflation at $N_{e} < 1$.  Thus, during inflation one can safely ignore the term $ \frac{2}{3\alpha}e^{2\sqrt{\frac{2}{3\alpha}}\vp} {\partial V\over \partial \theta}$ in the equations \rf{eqtheta} and \rf{Es2a} for $\theta$.  After the end of inflation, the coefficient  $ \frac{2}{3\alpha}e^{2\sqrt{\frac{2}{3\alpha}}\vp}$ becomes $O(1)$, so the evolution of the fields $\vp$ and $\theta$ shortly after the end of inflation can be well understood ignoring hyperbolicity of space-time.

\section{Inflation in $SL(2,\mathbb{Z})$ invariant models \label{Sec: axions}}   
Our cosmological models belong to a class of models where the field space  has a curved geometry
\begin{equation}
    \mathcal{S}=\int \mathrm{d}^4x \sqrt{-g} \left[-\frac{1}{2} \mathcal{G}_{IJ}\left(\phi^K \right)\partial_\mu \phi^I \partial^\mu \phi^J -V\left(\phi^K \right)\right] \;,
\label{S}\end{equation}
where $\mathcal{G}_{IJ}\left(\phi^K\right)$ is the metric on the field space. 
Equations of motion are
\be
\mathcal{D}_t \dot \phi^I + 3H \dot \phi^I + V^{,I}=0 \ ,
\label{eom}\ee
where $\mathcal{D}_t \phi^I=\dot{\phi^I} +\Gamma^I_{JK} \dot\phi^{J} \phi^K$ .

In hyperbolic geometry with $\tau= \theta+ie^{\sqrt{2\over 3\alpha} \varphi}$ and $\phi^1=\vp$ and $\phi^2=\theta$, $G_{11} =1, \, G_{22} =  {3\a\over 2} e^{-2\sqrt{\frac{2}{3\alpha}}\vp}$.
The action of the scalar fields in hyperbolic geometry is given by\footnote{Inflation and quantum fluctuations in the context of very similar models were studied in detail in \cite{Achucarro:2017ing,Linde:2018hmx}. In these models the metric of the $\theta$ field at large $\vp$ is of the form $\sim e^{2\sqrt{\frac{2}{3\alpha}}\vp}$} 
\begin{equation}\label{act}
\int d^4x\sqrt{-g}\left(-\frac12\partial^\mu\vp\partial_\mu\vp-\frac{3\alpha}{4}e^{-2\sqrt{\frac{2}{3\alpha}}\vp}\partial_\mu\theta\partial^\mu\theta-V(\vp,\theta)\right) \ .
\end{equation}
Equations of motion \rf{eom} for the homogeneous fields $\vp$ and $\theta$  are
\begin{equation}\label{eqphi}
\ddot\vp+3H\dot\vp+\sqrt{\frac{3\alpha}{2}}e^{-2\sqrt{\frac{2}{3\alpha}}\vp}\, \dot\theta^{2}+V_\vp=0 \ ,
\end{equation}
\begin{equation}\label{eqtheta}
\ddot\theta+3H\dot\theta-2\sqrt{\frac2{3\alpha}}\dot\vp\dot\theta+\frac{2}{3\alpha}e^{2\sqrt{\frac{2}{3\alpha}}\vp}V_\theta=0 \ .
\end{equation}
These two equations at $\vp \gg \sqrt{\alpha}$ in the slow-roll approximation are
\begin{equation}\label{Es1a}
3H\dot\vp = -V_\vp  \ ,
\end{equation}
\begin{equation}\label{Es2a}
3H\dot\theta= -\frac{2}{3\alpha}e^{2\sqrt{\frac{2}{3\alpha}}\vp}V_\theta\ .
\end{equation}
Note that although $V_\theta$ is small, it is multiplied by $e^{2\sqrt{\frac{2}{3\alpha}}\vp}$, which is very large during inflation. This could destabilize the trajectory with $\dot \theta =0$. 

However, in all  $SL(2,\mathbb{Z})$ invariant potentials presented in  \cite{Kallosh:2024ymt}, the plateau of the potentials at $\sqrt{\frac{2}{3\alpha}}\vp \gg 1$ is very flat not only with respect to the inflaton field $\vp$, which is a general property of all $\alpha$-attractors, but also with respect to the axion field $\theta$. Moreover, the smallness of $V_\theta$ is quite extraordinary, which can be explained by the fact that the  $SL(2,\mathbb{Z})$ invariants at large $\vp$ behave as double exponents in $\vp$, see eq. \rf{j2}.

Let us return to equations \rf{eqphi} and \rf{eqtheta} with this new understanding of the properties of the axion potential. The results obtained above imply that we can simply ignore  the last term in  \rf{eqtheta} during inflation, and this equation becomes
\begin{equation}\label{eqthetalight}
\ddot\theta+ 3H \dot\theta \ \bigl(1 -{2\over 3}\sqrt{ 2\over 3\alpha}{\dot\vp\over H}\bigr)=0 \ .
\end{equation}
In the slow-roll regime with $\dot\varphi <0$, this equation can be written as
 \begin{equation}\label{eqthetalight2}
\ddot\theta+ 3H \dot\theta \ \bigl(1 +{4\over 3}\sqrt{ \epsilon\over 3\alpha}\bigr)=0 \ ,
\end{equation}
where $\epsilon = {1\over 2} {V_{\vp}^{2}\over V^{2}} \ll 1$. In $SL(2,\mathbb{Z})$ invariant models, the  values of $3\alpha$ are discrete, $1, 2, 3,...,7$, see \cite{Kallosh:2024ymt}.  In such models, the last term in \rf{eqthetalight2} can be neglected, the Hubble constant at the beginning of inflation is approximately constant, and the solution  of this equation is given by
\be\label{thetaH}
\theta = \theta_{0}(1- e^{-3Ht}).
\ee
This implies that during the early stages of inflation with $H \approx const$, the speed of the field $\theta$ decreases exponentially,   $\dot \theta \propto e^{-3Ht}$, and within several e-foldings the field $\theta$ freezes,
\be
\theta = \theta_{0} \ .
\ee
Equation \rf{eqphi} with $\dot \theta = 0$ becomes the standard equation for the single-field $\alpha$-attractors, decoupled from the field $\theta$.
\begin{equation}\label{eqphi1}
\ddot\vp+3H\dot\vp +V_\vp=0 \ .
\end{equation}
Therefore, one may expect that the general inflationary predictions of the new class of models in the models with $\dot \theta = 0$ coincide with general predictions of single-field $\alpha$-attractors \rf{pred2}, see  \cite{Kallosh:2013yoa,Galante:2014ifa,Kallosh:2021mnu}. However, one must check whether the perturbations of the ultra-light axion field may lead to significant isocurvature perturbations, which could destabilize the inflationary trajectory and alter the amplitude of the curvature perturbations. 

Similar issues were recently discussed in  \cite{Cicoli:2008gp,Cicoli:2021yhb,Cicoli:2021itv} in the context of Fibre inflation derived in perturbative string theory. The potential of the inflaton field $\vp$ in Fibre inflation during the last 50-60 e-foldings nearly coincides with the $\alpha$-attractor potentials with $\alpha = 2$ \cite{Kallosh:2017wku}. The Fibre model of perturbative string theory was extended with non-perturbative terms of the type  $e^{-a T}$, which resulted in double-exponentially flat axion directions of the extended potential, similar to what we found in our $SL(2,\mathbb{Z})$ invariant models. The investigation performed in \cite{Cicoli:2021yhb,Cicoli:2021itv} shows that the isocurvature perturbations in their models do not affect their cosmological predictions. Since the models studied in \cite{Cicoli:2021yhb,Cicoli:2021itv} are somewhat similar to our models with $SL(2,\mathbb{Z})$ invariant potentials, one may expect that their conclusions regarding adiabatic and isocurvature perturbations apply to our models as well. 

However, the results obtained in  \cite{Cicoli:2021yhb,Cicoli:2021itv}  were based in part on the numerical investigation of the model with substantially different parameters, and the model itself was recently modified \cite{Cicoli:2024bxw}. In the following sections, we will perform an independent investigation of these issues in the context of our models with $SL(2,\mathbb{Z})$ invariant potentials.

\section{Background invariant action for linear perturbations  }\label{Sec:axions}

Following \cite{Iacconi:2021ltm}, we will investigate here the cosmological models \rf{S}.
 The equation of motion for the Fourier modes of scalar field perturbations {\it on flat
hypersurfaces} $Q^I= \delta \phi^I$ was derived in \cite{Sasaki:1995aw}
\begin{equation}
   \mathcal{D}_t\mathcal{D}_t Q^I+3H \mathcal{D}_t Q^I+\frac{k^2}{a^2} Q^i+{\mathcal{M}^I}_J Q^J=0 \ ,    
\label{field}\end{equation}
where
\begin{equation}
   {\mathcal{M}^I}_J\equiv {V_{;}^I}_J-{\mathcal{R}^I}_{KLJ}\dot{\phi}^K \dot{\phi}^L-\frac{1}{a^3}\mathcal{D}_t \left(\frac{a^3}{H}\dot{\phi}^I\dot{\phi}_J \right) \ .
\label{calM}\end{equation}
Here the covariant in field-space geometry ``long'' derivatives involve Christoffel  symbols due to the non-trivial geometry of the field-space
\be
\mathcal{D}_t A^I=\dot{A^I} +\Gamma^I_{JK} \dot\phi^{J} A^K \ ,
\label{long1}\ee  
\be
V_{;IJ}\equiv V_{,IJ}-\Gamma^{K}_{IJ}V_{,K} \ .
\label{long2}\ee
  The second term in eq. \rf{calM} depends on Riemann tensor ${\mathcal{R}^I}_{KLJ}$. For a two-dimensional field space, $I=1,2$ 
 the only independent component of 
the Riemann tensor is $\mathcal{R}_{IJKL}=\frac{1}{2}\mathcal{R}_{\text{fs}}\left(\mathcal{G}_{IK}\mathcal{G}_{JL}-\mathcal{G}_{IL}\mathcal{G}_{JK} \right)$, where $\mathcal{R}_{\text{fs}}$ is the intrinsic scalar curvature of the field space. In our case of hyperbolic geometry in eq. \rf{hyper} $\mathcal{R}_{\text{fs}}= -{4\over 3\a}$ (note that the supergravity \K curvature is $\mathcal{R}_{K}= -{2\over 3\a}$).
The third term in eq. \rf{calM} presents the gravitational backreaction due to spacetime metric perturbations induced by the field fluctuations at first order. 

If one prefers to work with gauge-invariant perturbations, one has to replace $Q^I= \delta \phi^I$ as follows
\be 
Q^I= \delta\phi^I - {\cal R}   {\dot \phi^I\over H} \ .
\label{R}\ee 
Here ${\cal R}$  is the intrinsic curvature perturbation of the constant time hypersurfaces, defined in \cite{Sasaki:1995aw}. The variables \rf{R} represent a multi-field version of the Mukhanov-Sasaki variables \cite{Mukhanov:1985rz,Sasaki:1986hm,Mukhanov:1988jd}. They are described, for example, in  \cite{Peterson:2010np}.

To study cosmology with ultra-light axions in FRW background, it is important to take into account an identity derived in \cite{Langlois:2008mn} for the second order action for perturbations $Q^I$ in multi-field inflation models with kinetic terms described by a non-trivial metric $G_{IJ} (\phi_K)$ in field space\footnote{In \cite{Langlois:2008mn} the action depends on a general function $P(X, \phi^I)$ where $X=-\frac{1}{2} \mathcal{G}_{IJ}\left(\phi^K \right)\partial_\mu \phi^I \partial^\mu \phi^J$. In that way, it covers cosmological models with higher derivatives, like the Dirac-Born-Infeld inflationary model.}.

In the modes which we study,   without higher derivatives $P(X, \phi^I)= X  -V(\phi^K)$, the 
 identity proven in \cite{Langlois:2008mn} takes a simple form, which we will present below and apply to our cosmological models with hyperbolic geometry.

The identity in our case states that the manifestly covariant second-order action $S_{(2)}$, where all derivatives are ``long'' derivatives as in eqs. \rf{long1}, \rf{long2},  is equal (up to total derivatives) to the action $S_{(2)}^{non-cov}$ where derivatives are ``short''.  This action $S_{(2)}^{non-cov}$  is not manifestly invariant, but it is equal to a manifestly invariant action $S_{(2)}$.  The identity is
\be
S_{(2)}^{non-cov}= S_{(2)}  \ .
\label{id}\ee
Here the manifestly invariant action is
\bea
S_{(2)}&=& {1\over 2} \int dt\, d^3x\, a^3\, [ G_{IJ} (\mathcal{D}_t Q^I \mathcal{D}_t Q^J -{1\over a^2 } \partial_i Q^I \partial^i   Q^J) - {\mathcal{M}}_{IJ} Q^I Q^J] \ ,
\label{cov}\eea
where ${\mathcal{M}}_{IJ} $  is related to the mass term in eq. \rf{calM} as follows
\begin{equation}
   {\mathcal{M}}_{IJ}= G_{IK}   {\mathcal{M}^K}_J \equiv {V_{;}}_{IJ}-{\mathcal{R}}_{IKLJ}\dot{\phi}^K \dot{\phi}^L-\frac{1}{a^3}\mathcal{D}_t \left(\frac{a^3}{H}\dot{\phi}_I\dot{\phi}_J \right) \ .
\label{calM1}\end{equation}
 Note that the space derivatives $\partial_i$ are background covariant in the FRW metric. Thus, each term in this action is background invariant.

The action \rf{cov} can be brought to the form $S_{(2)}^{non-cov}$ in which it is not invariant term by term; only the sum of all terms is invariant due to the Langlois-Renaux-Petel identity \rf{id}.
\be
S_{(2)}^{non-cov}= {1\over 2} \int dt\, d^3x\, a^3\,  [ G_{IJ} (\dot  Q^I \dot  Q^J -{1\over a^2 } \partial_i Q^I \partial^i   Q^J)+ 2G_{ IJ, K }\dot \phi^I  Q^K \dot Q^J  - (\mathbb{M}^2)_{IJ}  Q^I Q^J]\label{S2} \ .
\ee
Here
\be
(\mathbb{M}^2)_{IJ} = V_{, IJ}- {1\over 2} G_{KL, IJ }\dot \phi^K \dot \phi^L +(3-\epsilon) \dot \phi_I \dot \phi_J  +{1\over H} (V_{,I} \dot \phi_J + \dot \phi_I V_{,J})   \ .
\label{Mass2}
\ee
Here we used the fact that 
\be
-\frac{1}{a^3}\mathcal{D}_t \left(\frac{a^3}{H}\dot{\phi}_I\dot{\phi}_J \right)=(3-\epsilon) \dot \phi_I \dot \phi_J  +{1\over H} (V_{,I} \dot \phi_J + \dot \phi_I V_{,J})  \ .
\label{3d}\ee
 To prove \rf{3d}, we have used  the background field equations
\be
\mathcal{D}_t \dot \phi^I + 3H \dot \phi^I + V^{,I}=0\, , \qquad 
a^{-3} \mathcal{D}_t (a^3 \dot \phi^I) +V^{,I}=0 \ .
\ee
We may try to compare the action  \rf{S2} for linear scalar perturbations in the FRW  background with the quadratic part of the classical action for scalar fields in the FRW  background. The action \rf{S2} can be given as 
\bea \label{S2rel1}
&S_{(2)}^{Q}= {1\over 2} \int dt\, d^3x\, a^3 \,\Big ( - G_{IJ} \partial_\mu  Q^I \partial^\mu  Q^J   - V_{,IJ} (\phi) Q^I Q^J+ \\
\cr
&2G_{ IJ, K }\dot \phi^I  Q^K \dot Q^JV_{, IJ}-\Big [ {1\over 2} G_{KL, IJ }\dot \phi^K \dot \phi^L -(3-\epsilon) \dot \phi_I \dot \phi_J  -{1\over H} (V_{,I} \dot \phi_J + \dot \phi_I V_{,J}) \Big ] Q^I Q^J \Big ) \ .
\label{S2rel2} \eea
This is to be compared with the quadratic part of the classical action in eq. \rf{S}
\begin{equation}
    \mathcal{S}_{(2)}^{\rm cl}= \frac{1}{2} \int \mathrm{d}^4x \sqrt{-g} \left[- \mathcal{G}_{IJ} \partial_\mu \phi^I \partial^\mu \phi^J -V_{IJ} \phi^I \phi^J \right ]  \ .
\label{S2cl}\end{equation}
In general metric $G_{IJ}(\phi)$ the actions for perturbations $S_{(2)}^Q$ and classical action $ \mathcal{S}_{(2)}^{\rm cl}$ differ significantly, e.g. all terms in the second line in eq. \rf{S2rel2} are absent in \rf{S2cl}.

 We will now show that in $SL(2,\mathbb{Z})$ models with hyperbolic geometry and ultra-light axions, the action for the perturbations $Q^\theta\sim \delta \theta$ at small momentum $k\to 0$  is the same as the action for the homogeneous classical field $\theta$.

\section{ $SL(2,\mathbb{Z})$ models  with hyperbolic geometry and ultra-light axion}  \label{ultra}

Consider the gauge-invariant perturbations in eq. \rf{R}. The term correcting $\delta \phi^I$ to gauge-covariant $Q^I$ depends on 
 the intrinsic curvature perturbation of the constant time hypersurfaces ${\cal R}$, but it is also proportional to velocity $\dot \phi^I$.
If  a consistent background solution in some direction $\hat I$  with $\dot \phi^{\hat I}=0$ exists we find that during inflation
\be
Q^{\hat I} |_{\dot \phi^{\hat I}=0} = \delta\phi^{\hat I} \ .
\ee
It  means that even in a gauge where 
 ${\cal R}\neq 0$,  in some direction $\hat I$ where $\dot \phi^{\hat I}=0$
the gauge-invariant perturbation is just a scalar field perturbation.

This is precisely the case of $SL(2,\mathbb{Z})$ models with hyperbolic geometry and ultra-light axion $\theta$, where during inflation we have an ultra-light background field $\theta$ with vanishing background velocity and vanishing derivatives of the potential.
\be
\dot \theta=0\, ,\qquad V_{, \theta}=0 \, ,\qquad V_{, \theta \vp}=0  \, ,\qquad V_{, \theta \theta}=0 \ .
\ee

We will now apply systematically the study of perturbations above for general metric $G_{IJ}(\phi) $ and arbitrary potentials to the case of 
 $SL(2,\mathbb{Z})$ cosmological models in FRW background.  Our fields are $\phi^1=\vp, \, \phi^2= \theta$, and $G_{11}=1, G_{12}=0, G_{22} =  {3\a\over 2} e^{-2\sqrt{\frac{2}{3\alpha}}\vp}$, see eq. \rf{act}. In such case, 
the action for inflationary perturbations  takes the form
\be
S_{(2)}= {1\over 2} \int d^4 x a^3 [ G_{IJ} (\dot  Q^I \dot  Q^J -{1\over a^2 } \partial_i Q^I \partial^i   Q^J) + 2G_{ \theta \theta, \vp }\dot \theta  Q^\vp \dot Q^\theta  - (\mathbb{M}^2)_{IJ}  Q^I Q^J]  \ .
\ee
Here
\be
(\mathbb{M}^2)_{IJ} = V_{, IJ}- {1\over 2} G_{\theta \theta, IJ }\dot \theta \dot \theta +(3-\epsilon) \dot \phi_I \dot \phi_J  +{1\over H}(V_{,I} \dot \phi_J + \dot \phi_I V_{,J})  \ .
\ee
During inflation, the expression for the mass matrix simplifies since
 $\dot \theta= {\partial V\over \partial\theta}=  {\partial^2 V\over \partial\theta \partial \vp} =  {\partial^2 V\over \partial\theta \partial \theta} =0 $. Therefore
\be
 (\mathbb{M}^2)_{IJ}^{\rm inf l}=\left(\begin{array}{cc} (\mathbb{M}^2)_{\vp \vp} & 0 \\0 & 0\end{array}\right) \ .
\label{mass2}\ee 
This result confirms the related result obtained in \cite{Cicoli:2021yhb} by a different method.

 Thus, perturbations in the inflationary background have the following action
\be\label{s2}
S_{(2)}= {1\over 2} \int d^4 x a^3 [   (\dot  Q^\vp \dot  Q^\vp-{1\over a^2 } \partial_i Q^\vp \partial^i   Q^\vp) + G_{\theta\theta} (\dot  Q^\theta \dot  Q^\theta-{1\over a^2 } \partial_i Q^\theta \partial^i   Q^\theta)
 + (\mathbb{M}^2)_{\vp \vp}  Q^\vp Q^\vp] \ .
\ee
The action for  $  Q^\theta$ perturbations is decoupled from the action for $  Q^\vp$ perturbations and,  for inflationary perturbations with exponentially large wavelengths,  has the same form as the action for the background field $\theta$ 
\be
 {1\over 2} \int d^4 x a^3 \, G_{\theta\theta} \dot  Q^\theta \dot  Q^\theta   \qquad \Leftrightarrow   \qquad {1\over 2} \int d^4 x a^3 \,  G_{\theta\theta} \dot  \theta \dot  \theta  
\ee
The action for $  Q^\vp$ perturbations even for $k\to 0$ is different from the action for a background scalar $\vp$ since the last term in the expression
\be
(\mathbb{M}^2)_{\vp \vp} = V_{, \vp \vp} -\frac{1}{a^3}\mathcal{D}_t \left(\frac{a^3}{H}\dot \vp \dot \vp \right)  \ .
\ee
is due to spacetime metric perturbations. This term is not present in the equation for the background field $\vp$.

It is important that the action \rf{s2}  does not have any terms depending simultaneously on $Q^\vp$ and $Q^\theta$. Therefore we have two independent equations for  $Q^\vp$ and $Q^\theta$ in the background $\dot \theta = 0$: 
\begin{align}
\label{adiabatic eq}
    \Ddot{{Q}}^\vp +3H \dot{{Q}}^\vp + \left(\frac{k^2}{a^2} +{\mathcal{M}}^{\vp}{}_{\vp} \right){Q}^\vp &= 0\;, \\ 
\label{isocurvature}    
    \Ddot{{Q}}^\theta +(3H + 2\Gamma^\theta_{\theta \vp} \dot \vp) \dot{{Q}}^\theta + \left(\frac{k^2}{a^2}  \right){Q}^\theta &= 0 \;.
\end{align}
The first of these two equations is the standard equation for adiabatic perturbations in the single-field inflation models; its solutions are well-known. 
Therefore, all curvature perturbations produced at that stage with $\dot \theta = 0$ coincide with the perturbations in the single-field $\alpha$-attractors. 

The second equation can be further simplified. First of all, during inflation with $\dot\theta = 0$, the second term in the expression $Q^\theta= \delta \theta - {\cal R} {\dot \theta^i\over H}$ vanishes, so the amplitude of isocurvature perturbations $Q^\theta$ is equal to the amplitude of the fluctuations $\delta \theta$.

Also,  $\Gamma^\theta_{\theta \vp}= -\sqrt{2/3\a}$, which means that  $
3H + 2 \Gamma^\theta_{\theta \vp} \dot \vp = 3H \Big (1-2\sqrt{2\over 3\a}\, {\dot \vp\over 3H}\Big)$.
The full equation for   perturbations $Q^\theta= \delta\theta$ is 
 \begin{equation}\label{eqthetalightdelta1}
\ddot{\delta}\theta+ 3H \dot{\delta \theta} \ \bigl(1 -{2\over 3}\sqrt{ 2\over 3\alpha}{\dot\vp\over H}\bigr)+\left(\frac{k^2}{a^2}  \right)\delta \theta= 0 \ .
\end{equation}
In the limit $k/a \ll H$, i.e., for the modes with wavelengths greater than the horizon $\sim H^{{-1}} $, one can neglect the last term, and the equation for the perturbations $\delta\theta$ becomes the same as the equation for the homogeneous field $\theta$ \rf{eqthetalight}.

In the slow-roll regime with $\alpha = O(1)$ one has  $-{2\over 3}\sqrt{ 2\over 3\alpha}{\dot\vp\over H}  = {4\over 3}\sqrt{ \epsilon\over 3\alpha}\ll 1$,   and the equation \rf{eqthetalightdelta1}  becomes
\be\label{QQQ}
 \Ddot{\delta\theta} +3H  \dot{\delta \theta} + \left(\frac{k^2}{a^2}  \right)\delta \theta = 0 \ .
\ee
This equation coincides with equation (7.3.5) from the book  \cite{Linde:1990flp}, so we can use its solution presented there in equation (7.3.8).  However, one should take into account that the field $\delta \theta$ is not canonically normalized, its metric \rf{act} contains an additional coefficient  $\frac{3\alpha}{2}e^{-2\sqrt{\frac{2}{3\alpha}}\vp}$.  During the quantization of high-frequency fluctuations of the field $\theta$ with $k > H$, one can ignore the slow evolution of the field $\vp$ and simply rescale the field $\theta$ by the factor $\sqrt{\frac{3\alpha}{2}}e^{-\sqrt{\frac{2}{3\alpha}}\vp}$ to make it canonical. 

This implies that the amplitude of quantum fluctuations of the field $\theta$ is greater than the amplitude of fluctuations of the canonically normalized field given in \cite{Linde:1990flp} by the factor $\sqrt{\frac{2}{3\alpha}}e^{\sqrt{\frac{2}{3\alpha}}\vp}$. Comparing it with equation (7.3.8) of    \cite{Linde:1990flp} finally yields\footnote{We are grateful to R. Gonzalez Quaglia,  M. Michelotti, and  D. Roest for the discussion of this normalization factor. A similar factor was previously discussed in our paper on the multi-field $\alpha$-attractors \cite{Achucarro:2017ing}, but the exponent in those models had a different sign \cite{Kallosh:2024ymt}, which resulted in the exponential suppression of the amplitude of the perturbations $\delta \theta$. In the context of the $SL(2,\mathbb{Z})$ invariant $\alpha$-attractors, the amplitude of the perturbations $\delta \theta$ is exponentially enhanced. }
\be
\label{7.3.8}
\delta\theta_k(t)= \sqrt{\frac{2}{3\alpha}}e^{\sqrt{\frac{2}{3\alpha}}\vp}\ \frac{i\,{ H}}{k\,\sqrt{2\,k}}\,
\left(1+\frac{k}{i\,{\rm H}}e^{-{ H}\,t}\right)\,
\exp\left(\frac{i\,k}{{\rm H}}\,e^{-{ H}\,t}\right)\ .
\ee
At large $t$, when $k\,e^{-{  H}\,t}$ becomes smaller than $H$, which corresponds to the horizon crossing, the solution ceases to oscillate and becomes equal to
\be
\delta\theta_k(t)|_{k\,e^{-{H}\,t}<{ H}}\quad  \to \quad \displaystyle \frac{i\,{\rm H}}{k\,\sqrt{2\,k}} \, \sqrt{\frac{2}{3\alpha}}e^{\sqrt{\frac{2}{3\alpha}}\vp}\  .
\label{Qtheta}\ee
Eq. (7.3.12) in \cite{Linde:1990flp} explains that this leads to the well-known expression for the amplitude of the perturbations generated during a single e-folding:
\be
\delta \theta ={H\over 2\pi} \sqrt{\frac{2}{3\alpha}}e^{\sqrt{\frac{2}{3\alpha}}\vp} \ .
\label{delta}\ee
Using the relation \rf{ne}, one can represent this result in a different form:
\be
\delta \theta \approx {H\over 2\pi} \,  \left(\frac{2}{3\alpha}\right)^{3/2} 4 N_{e} \ .
\label{Ndelta}\ee
The extra factor $\left(\frac{2}{3\alpha}\right)^{3/2} 4 N_{e}$. can be quite significant. For example, for $\alpha = 1/3$, $N_{e} = 50$ one has
\be\label{Ndelta13}
\delta \theta \approx 8\sqrt{2} N_{e}\, {H\over 2\pi}  \approx   566\   {H\over 2\pi}  \ .
\ee

Thus we find that during each e-folding of inflation with $\dot \theta = 0$, the isocurvature perturbations
\be
Q^\theta=  \delta\theta  - {\cal R} {\dot \theta \over H}
=  \delta\theta 
\ee
freeze at a constant value \rf{delta}, \rf{Ndelta}, \rf{Ndelta13} when the wavelengths of these perturbations become  greater than $H^{{-1}}$.

One may wonder whether these perturbations may ``unfreeze'' and begin to grow at the very end of inflation when the field $\vp$ continues rolling down, and the slow-roll approximation is no longer valid. This could happen if the friction coefficient $3H \Big (1-\sqrt{\a\over 6}\, {\dot \vp\over H}\Big)$ in equation  \rf{isocurvature} changes its sign. But $\dot \vp < 0$ when the field $\phi$ rolls down. Therefore, the friction coefficient is even greater than $3H$ for $\dot\vp < 0$. This leads to a strong stabilization of the inflationary trajectory $\theta = const$ and of the superhorizon perturbations $\delta \theta$.

 These considerations address the issue of stability of the inflationary trajectory $\dot \theta = 0$: during inflation and shortly after its ends, the classical trajectory  $\dot \theta  = 0$ of the ultra-light axion field is stable, as long as one can ignore the double-exponentially suppressed derivatives $V_{\theta}$ and $V_{\theta,\theta}$.
 
 However, there are small inflationary perturbations  \rf{delta} near each classical trajectory $\theta = const$. At the end of inflation, the trajectory may turn towards one of the minima of the potential. That is the time when isocurvature perturbations could feed into the adiabatic ones. We will discuss this possibility in the next section.

\section{Adiabatic perturbations from the isocurvature ones?}\label{Sec:adiab}

As we already mentioned,  the field $\theta$ remains constant during inflation in our models. In this sense, such models are similar to the single-field inflationary models, where the amplitude of curvature perturbations  on comoving hypersurfaces ${\cal{R}}_c$  has a simple interpretation in the $\delta N$ approach   \cite{Sasaki:1995aw, Sasaki:1998ug}: 
\be \label{ad}
{\cal{R}}_c = \delta N_{\vp} = H \delta t = H{\delta\vp \over \dot\vp} =  {H^{2} \over 2\pi \dot \vp} \ .
\ee
Here    $\delta N_{\vp}$ is the change of the number of e-foldings due to the fluctuations with the average amplitude $|\delta\vp|  = {H\over 2\pi}$, which may either increase or decrease the number of e-folds until the end of inflation, depending on the sign of $\delta\vp$. All quantities here should be evaluated at the moment $N_{e}$ e-foldings before the end of inflation. To compare the results with the CMB data, one should consider $N_{e} \sim 50.$

The amplitude $A_{s}$ usually presented in CMB data is given by the square of this quantity, 
\be\label{A}
A_{s} =  \delta N_{\vp}^{2} = \left({H^{2} \over 2\pi \dot \vp}\right)^{2}   \ .
\ee
For $\alpha$-attractors it is given by \cite{Kallosh:2013yoa,Galante:2014ifa,Carrasco:2015uma,Kallosh:2021mnu}
\be\label{B}
A_{s} =  {V_{0}\, N_{e}^{2}\over 18 \pi^{2 }\alpha} \ .
\ee

A comparison of \rf{A} and \rf{B} shows that the amplitude of perturbations $\delta N_{\vp}$ in these models is directly proportional to $N_{e}$. Since the Hubble constant does not change much until the end of inflation in these models, these equations imply that $\dot\vp$ is inversely proportional to $N_{e}$. This means, for example, that the velocity of the field $\vp$ $50$ e-foldings prior to the end of inflation is approximately $10$ times smaller than the velocity of the field $\vp$ at $N_{e} = 5$, and O(50) times smaller than the velocity of the field $\vp$ at $N_{e} = 1$.  This is a rough estimate, but the result is qualitatively correct, especially for $\alpha <1$: The value of $\dot\vp$ at the end of inflation is significantly greater than the value of $\dot\vp$ at the moment corresponding to $N_{e} \sim 50$.

As long as the field trajectory remains straight with $\theta = const$, the perturbations of the field $\theta$ do not induce any perturbations $\delta N_{\theta}$ because they are orthogonal to the field trajectory.  If, for example, the field $\theta = \theta_{0}$ during inflation was equal to $0.5$ (or, more generally $0.5+k$, where $k$ is an integer), then this field trajectory never changes its direction even after inflation ends, because the potential has a minimum along the direction $\theta = 0.5$. This trajectory is stable, $V_{\theta} (\pm 0.5)  = 0$ and  $V_{\theta,\theta} (\pm 0.5)> 0$ for all values of the field $\vp$ along the straight trajectory, which brings the field $\vp$ towards the minimum of the potential at  $\theta = 0.5$, $\vp_{1}= \sqrt{3\alpha\over 2} \ln{\sqrt 3\over 2}$.  In that case (no turns of the field trajectory), the perturbations $\delta \theta$ do not affect the amplitude of curvature perturbations \rf{ad}, and the cosmological predictions coincide with the standard predictions of the single-field $\alpha$-attractors.

However, any trajectory with $\theta_{0} \not = 0.5 + k$ eventually turns towards one of the minima of the potential at $\vp <0$. At that time, some of the perturbations $\delta \theta$ generated at early stages of inflation may feed into the curvature perturbations.

To make a rough estimate of the curvature perturbations that may be generated due to the turn, we consider an inflationary trajectory with $0< \theta_{0}< 0.5$. Suppose that the field $\vp$ continues moving along this trajectory until it reaches $\vp_{1}= \sqrt{3\alpha\over 2} \ln{\sqrt 3\over 2}$ and them rapidly turns in $\theta$ direction towards the minimum at  $\theta = 0.5$, $\vp_{1}$.  At the beginning of the turn, the field was not exactly at $\theta = \theta_{0}$, but at $\theta = \theta_{0} \pm \delta\theta$. The parts of the universe that were closer to the minimum of the potential by $\delta\theta$ will reach it earlier by $\delta t =  \delta\theta/\dot\theta$. The perturbations $\delta\theta$  formed  at $N_{e} = 50$  have an initial magnitude given by \rf{delta}-\rf{Ndelta13}. 

If the turn occurs during inflation, using equation \rf{Ndelta}, one finds that such perturbations may lead to a change in the number of e-foldings by
\be \label{isoad}
\delta N_{\theta} \sim H \delta t = H{\delta\theta\over \dot\theta} \approx   4 N_{e} \left(\frac{2}{3\alpha}\right)^{3/2}  {H^{2}\over 2\pi\dot \theta}  \ \  .
\ee
These perturbations are adiabatic, and they are added to the previously generated adiabatic perturbations  \rf{ad} with the same large wavelength $\sim e^{N_{e}}$.

Strictly speaking, one should not use this equation after inflation. However, we expect that the local delay of the onset of reheating due to the fluctuations  $\delta\theta$ may lead to a qualitatively similar effect. 

To complete this estimate, we would need to know the value of $\dot \theta$ in \rf{isoad} shortly after the fields turn. If this turn happens very fast, the kinetic energy of the field $\theta$ after the turn nearly coincides with the kinetic energy of the field $\phi$ before the turn. The field $\theta$ is not canonically normalized, its kinetic energy is  $\frac{3\alpha}{4}e^{-2\sqrt{\frac{2}{3\alpha}}\vp}(\dot\theta)^{2}$, see \rf{act}. One can check that at $\vp = \vp_{1}$ this expression reduces to $\alpha (\dot\theta)^{2}$. Therefore, energy conservation tells that $\dot \theta \approx {\dot\vp\over \sqrt{2\alpha}}$, where $\dot \vp$ is the velocity of the field $\vp$ shortly before the turn, which is not much different from the velocity of this field at the end of inflation.

As we already argued, the value of   $\dot \vp$ at the end of inflation (and therefore the value of $\dot \theta$ after the turn, which enters equation \ref{isoad})  is about $N_{e}$ times greater than the value of $\dot \vp$ used in \rf{ad}, which is evaluated at the moment $N_{e}$ e-foldings before the end of inflation. Bringing all of these factors together, one can make the following estimate for the relation between $\delta N_{\theta}$ and $\delta N_{\vp}$:
\be \label{isoad2}
\delta N_{\theta} \sim  {16 \over 3\, \alpha \sqrt 3 }\ \delta N_{\vp}   \  .
\ee
In particular, for $\alpha = 1/3$ one has $\delta N_{\theta} \sim  10\, \delta N_{\vp}$. This suggests that the amplitude of the curvature perturbations $\delta N_{\theta}$ generated by the post-inflationary conversion of the isocurvature perturbations can be much greater than the standard amplitude $ \delta N_{\vp}$.  And the relative contribution of $\delta N_{\theta}$ to $A_{s}$ is even much greater.

Indeed,  perturbations $\delta\vp$ and $\delta\theta$ (and their sign) are not statistically correlated. Therefore, one may estimate the resulting value of $A_{s}$ as
\be
A_{s} \approx  \delta N_{\vp}^{2}+\delta N_{\theta}^{2} \ . 
\ee
Thus if $\delta N_{\theta}$ is 10 times greater than  $\delta N_{\vp}$, its relative contribution to  $A_{s}$ is 100 times greater than the standard single-field contribution $\delta N_{\vp}^{2}$. 

These order-of-magnitude estimates are very preliminary and imprecise. In particular, the results of a more detailed investigation should strongly depend on the initial value of $\theta$. For example, as we already mentioned in the previous section, the trajectory with $\theta = \pm 0.5$ never turns during and after inflation, so in that case $\dot \theta = 0$ even after inflation. For such initial values of $\theta$ (no bending), the isocurvature perturbations do not feed to the curvature perturbations, $\delta N_{\theta}= 0$, in which case the inflationary predictions of $SL(2,\mathbb{Z})$ invariant models should coincide with the predictions of the single-field $\alpha$-attractiors. 

This may suggest that the closer the initial value of the field $\theta$ is to the position of the minimum at $\theta = 0.5$, the smaller is the effect. Conversely, one may expect a large effect if the initial value of the field is close to $\theta = 0$, because then the trajectory must significantly turn and go a long way towards the minimum at $\theta = 0.5$.  However, the structure of the $SL(2,\mathbb{Z})$ invariant potentials at $\vp \lesssim 0$ is very complicated. 
Therefore, a much more detailed investigation is required to study this process, and the results should depend not only on the initial values of $\theta$  but also on the choice of the $SL(2,\mathbb{Z})$ invariant model. We will return to this question in a forthcoming publication \cite{6authors}. 

\section{Discussion}

In this paper, we explored the structure of the inflationary plateau of the $SL(2,\mathbb{Z})$ invariant $\alpha$-attractor potentials introduced in \cite{Kallosh:2024ymt,Casas:2024jbw}. We have found that the derivatives of these potentials in the axion direction $\theta$  are double-exponentially suppressed during inflation at large values of the inflaton field $\vp$. For example, during inflation the potential in eq. \rf{Renata2} is 
\be
V \to V_0\Big (1-c_1e^{-\sqrt{2\over 3\alpha} \varphi} +  c_2  \, e^{-2\sqrt{2\over 3\alpha} \varphi} e^{-2 \pi e^{\sqrt{2\over 3\alpha} \varphi}} \cos(2\pi \theta)\Big )\ .
\label{beta2}\ee
where the coefficients $c_1$ and $c_2$ depend on the choice of the modular potentials in  \cite{Kallosh:2024ymt,Casas:2024jbw}.

Because of the extraordinary flatness of the axion potential, the axion field freezes and does not evolve during inflation. Therefore in typical models of this type, the only evolving field is the inflaton. We developed a theory of curvature and isocurvature perturbations in these models and demonstrated that the isocurvature perturbations do not destabilize the inflationary trajectory. Thus, inflation in this theory is effectively single-field. 

Usually, this is where the investigation of the inflationary perturbations ends. However, as discussed in sections \ref{ultra} and \ref{Sec:adiab}, due to the hyperbolic geometry of the field space, the amplitude of the fluctuations $\delta \theta$ can be very large. Therefore it is important to check whether these perturbations may feed into the curvature perturbations at the post-inflationary epoch and affect the cosmological predictions. 

We have found that these perturbations do not affect the amplitude of the curvature perturbations for the inflationary trajectories with $\theta = \pm 0.5$. However, our preliminary estimates suggest that the perturbations of $\delta \theta$ may significantly affect the amplitude of the curvature perturbations for $\theta$ far away from $\pm 0.5$.

A more detailed investigation is required to obtain a complete and reliable expression of the amplitude of adiabatic perturbations in this scenario.  The corresponding results will depend on the initial values of $\theta$. Moreover, these results will be model-dependent because the post-inflationary evolution depends on the complicated and model-dependent structure of $SL(2,\mathbb{Z})$ invariant potentials close to their minima. The results of our numerical and analytical investigation of these issues will appear in a separate paper \cite{6authors}.

Finally, one should note that the simplest  $SL(2,\mathbb{Z})$ invariant potential \rf{Renata2} studied in our paper was a function of $|j(\tau)|^{2}$. That is why the $\theta$-dependent part of the potential \rf{beta2} is double-exponentially suppressed. However, the requirement of the $SL(2,\mathbb{Z})$ invariance allows us to consider many other potentials as well. Indeed, $j$-function  is $SL(2,\mathbb{Z})$ invariant, and therefore {\it any} function of $j$ and $\bar j$ is also  $SL(2,\mathbb{Z})$ invariant.  
The simplest possibility would be to replace $|j(\tau)|^{2}$ in the potential \rf{Renata2} by $|j(\tau)|^{2} + A | j(\tau) \pm \overline j(\tau)|^{2}$, where $A$ is some constant. Such modification leads to a strong axion stabilization and completely eliminates isocurvature perturbations, even for rather small values of the stabilization parameter $A$   \cite{Carrasco:2025rud}.  The cosmological predictions of such models with strong axion stabilization coincide with the standard predictions of the single-field $\alpha$-attractors \rf{pred2}.

\section*{Acknowledgement}
We are grateful to M. Braglia,  D. Wands, T. Wrase, and Y. Yamada for helpful discussions and especially to J.J. Carrasco, R. Gonzalez Quaglia,  M. Michelotti, and  D. Roest for many fruitful discussions of $SL(2,\mathbb{Z})$ cosmology and for the ongoing collaboration on inflation in these models.
Our work is supported by SITP and by the US National Science Foundation Grant   PHY-2310429.

\bibliographystyle{JHEP}
\bibliography{lindekalloshrefs}
\end{document}